\documentclass[prl,twocolumn,groupedaddress]{revtex4}
\usepackage{graphicx}
\usepackage[compact]{titlesec}
\usepackage{sidecap}
\usepackage[caption = false]{subfig}

\usepackage{floatrow}

\usepackage{calc}
\usepackage{pifont}
\usepackage{amssymb}
\usepackage{epsfig}
\usepackage{amssymb}
\usepackage{amsmath}
\usepackage{color}
\usepackage{braket}

\sidecaptionvpos{figure}{c}

\DeclareMathAlphabet{\mathitb}{OT1}{cmr}{bx}{sl}
\begin{document}

\renewcommand{\thefootnote}{\fnsymbol{footnote}}

\title{Controlled Electrode Magnetization Alignment in Planar Elliptical Ferromagnetic Break Junction Devices}

\author{Gavin D. Scott$^1$}
\email{gavin.scott@nokia.com}
\author{Ting-Chen Hu$^1$}
\affiliation{$^1$Bell Laboratories, Nokia, 600 Mountain Ave, Murray Hill, NJ 07974, USA}

\date{\today}

 \begin{abstract}

Controlling the magnetization reversal process of magnetic elements is important for a wide range of applications that make use of magnetoresistive effects, but is difficult to achieve for devices that require adjacent thin film structures capable of contacting an individual molecule or quantum dot.  We report on the fabrication and measurement of ferromagnetic break junction devices with planar, elliptical leads to address the particular challenge of controlling the relative magnetization alignment between neighboring electrodes.  Low temperature transport measurements, supported by finite-element micromagnetic simulations, are used to characterize the magnetoresistance response across a range of conductance levels.  We demonstrate that an in-plane external field applied parallel to the hard axis of the ellipses may be used to controllably switch the magnetization of the source and drain electrodes between monodomain-like parallel and antiparallel configurations for devices in the tunneling regime.

\end{abstract}

\maketitle

\section{INTRODUCTION}

Understanding the detailed mechanisms behind the magnetization reversal process of magnetic elements is of both fundamental and commercial interest.  Devices based on spin valves make use of magnetoresistance (MR) that occurs as the relative magnetization alignment between two or more stacked magnetic thin film layers, separated by non-magnetic materials, alternate between parallel (P) and antiparallel (AP) configurations.  Proposals have also been put forward for applications that will require coplanar magnetic films that can similarly be switched between P and AP configurations, while separated by only a small island or quantum dot, as in a single molecule transistor.\cite{Kulinich2014,Ilinskaya2015,Chen2008,Kirchner2005}  The utility of these proposals is predicated upon the ability to reduce or eliminate the local effective field felt by the island when the magnetic moments of the opposing sides possess an AP orientation.  We demonstrate a means of controlling the relative magnetization orientation, which, until now, has presented a barrier to any implementation of these applications.

A few experimental efforts have exploited electrode shape anisotropy to influence the pair-wise magnetization reversal process in ferromagnetic break junctions, and have achieved devices that may be switched from collinear (i.e. P) to non-collinear configurations,\cite{Pasupathy2004,Keane2006,Yoshida2013} or to a local AP configuration through the creation of neighboring vortex states\cite{Bolotin2006,Bolotin2006b,Shi2007}.  However, the homogeneous (monodomain) AP configuration, required for eliminating the inter-electrode field and ensuring maximum spin polarization, has proven difficult to achieve.  Definitively demonstrating such a P to AP transition is complicated by the fact that MR measurements alone are not solely indicative of what is happening at nanocontact tip region and because the magnetization of the leads and the nanocontact region at the break are not trivially related.

We have fabricated ferromagnetic break junction devices designed specifically to allow for tunable switching between P and AP configurations.  Each sample consists of a pair of closely-spaced planar ellipses connected by a narrow constriction.\cite{Scott2016}  The nanoscale contacts are formed by reducing the size of a constriction using an electromigration technique.  The reversal process is analyzed via MR measurements, which are shown to be consistent with numerical simulations.  Results are presented for external field orientations parallel to both the short(hard) and long(easy) axis of the ellipses.

At conductance regimes up to $R{\lesssim}100k\Omega$ the MR characteristics and relatively small values of MR ratio (defined here as MRR$~=~(R_{max}-R_{min})/R_{P}\times{100}$, where $R_P$, the resistance with P magnetization, is equal to $R_s$, the resistance in a saturating field) may be accounted for by spin-orbit scattering mechanisms influenced by the electrode shape anisotropy and dipole-dipole interaction.  These mechanisms include anisotropic MR (AMR), atomically-enhanced AMR (AAMR), and tunneling AMR (TAMR).  The central finding is that after a sufficient tunneling gap is established ($R>100k\Omega$) the magnetization of the electrodes may be controllably switched between a P configuration and a monodomain-like AP configuration only when the field is applied along the hard-axis of the ellipses.  In this transport regime MR features will be dominated by tunneling MR (TMR) due to spin-dependent transport across the vacuum gap, but may be partially counterbalanced by TAMR and AMR contributions, leading to MRR values that are not observed to exceed $60\%$.

\section{FABRICATION}

\begin{figure}
\begin{center}
\includegraphics[scale = .29]{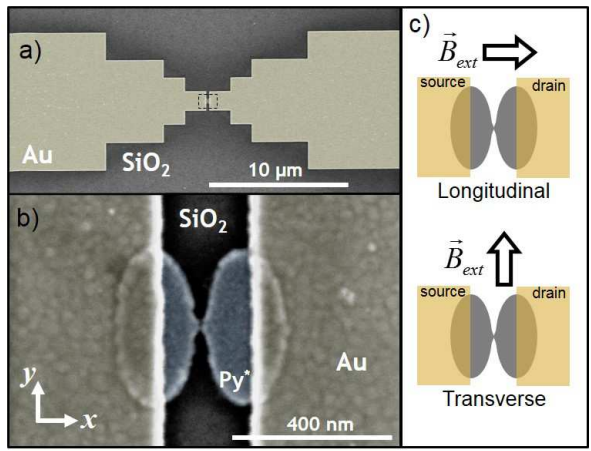}
\end{center}
\vspace{-5mm}
\caption{(a) False color scanning electron microscope image of Ti/Au electrodes overlapping opposing edges of the magnetic ellipse-pair.  (b) Zoom in of region marked by black dashed box in (a) showing the device after using electromigration to create a tunneling gap between the ellipses.  (c) An in-plane external magnetic field was applied using two configurations - longitudinal and transverse to the direction of current flow.}
\label{Fab1}
\end{figure}

Arrays of devices were defined by ebeam lithography on a highly doped $n$-type silicon substrate with a $2000${\AA} dry thermal oxide.  The ellipse-pairs were composed of a $15${\AA} Ti adhesion layer and $150${\AA} of supermalloy (Py$^*$:~Ni$_{80}$Fe$_{14}$Mo$_{5}$X$_{1}$ where X is other metals) deposited by ebeam evaporation.  This magnetic alloy was used due to its lack of magnetocrystalline anisotropy, low magnetostriction, and large spin polarization at the Fermi level, similar to permalloy (Py:~Ni$_{80}$Fe$_{20}$).  The ellipses have $200nm$ short-axis diameters and $400nm$ long-axis diameters aligned parallel to one another and perpendicular to the constriction between them (Fig.~\ref{Fab1}).  For a given pair, the center-to-center distance is $\sim230nm$, and the minimum lateral width of a constriction ranges from $50-100nm$.  Larger non-magnetic contacts were then defined by ebeam lithography in order to make electrical contact with the ellipses.  Oxides forming on the surface of Py$^*$ exposed to atmosphere are expected to contain a non-uniform depth profile akin to Py with a layer anatomy dependent upon factors including temperature, environmental oxygen concentration, and exposure time.\cite{Pollak1975,Bruckner2001,Fitzsimmons2006,Salou2008}  Depositing the magnetic devices on top of the larger electrodes\cite{Bolotin2006,Bolotin2006b,Shi2007} would circumvent complications associated with the formation of an oxide barrier between the two materials, but would have the consequence of creating a step in each ellipse where the overlap occurs, sacrificing the integrity of the desired shape anisotropy and accessible intermediate magnetization states.

Since the Py$^*$ ellipses are the bottom layer here, it is critical to remove a sufficient amount of the native oxide prior to the ensuing deposition of the Ti/Au ($50/1200${\AA}) electrodes in order to obtain an adequately low contact resistance.  The effectiveness of several etch recipes was ascertained by the aggregated room temperature 2-terminal resistance measurements through many ($>100$) individual devices after the final thin film layer was added (Ti/Au $50/1200${\AA}) using standard UV photolithography.  An alkaline wet etch solution of (NH$_4$OH:H$_2$O $1:20$ for $120s$) was found to work well.  We note that although the devices were exposed to atmosphere for 1 to 2 minutes while being transferred to the evaporation chamber, the oxidation process for Ni-Fe alloys has been shown to be less efficient, and therefore proceeds more slowly, than for pure Ni or Fe due to poorer penetration of the material by oxygen.\cite{Salou2008}

\section{MEASUREMENT AND SIMULATION DETAILS}

Thermally assisted electromigration was performed in vacuum at low temperature ($T{\lesssim}4K$) in order to incrementally reduce the size of a constriction until an atomic scale contact or tunneling gap was formed.\cite{Park1999}  Upon halting the electromigration process, differential conductance ($dI/dV$) was first measured as a function of both source-drain bias ($V_{sd}$) and gate bias ($V_G$) using a low frequency lock-in technique.  The silicon substrate was used as a global back gate.  Transport data presented here were acquired at $40mK\leq{T}\leq{2K}$.

Of the 51 clean Py$^*$ devices tested, $98\%$ exhibited only weakly nonlinear $dI/dV$ vs. $V_{sd}$ traces with no gate dependence or zero bias anomaly, consistent with clean nanoconstrictions and tunneling gaps in break junction experiments.  Experimental signatures of MR were acquired by measuring DC resistance at small bias voltages, to avoid heating effects, as the externally applied magnetic field, $\textbf{B}_{ext}$ was swept from a saturating field $\textbf{B}_s$ to $-\textbf{B}_s$ and then back.  $\textbf{B}_{ext}$ was aligned along the \emph{y} (easy) axis of the ellipses, transverse to the direction of current flow, for a portion of samples and along the \emph{x} (hard) axis of the ellipses, longitudinal to the direction of current flow, for the remaining samples.

The magnetization reversal process in a pair of closely spaced ferromagnetic planar ellipses was analyzed using the finite-element micromagnetic simulation package $Nmag$,\cite{Nmag} which integrates the Landau-Lifshitz-Gilbert equation numerically over time\cite{Nmag,Wei2007} at incrementally changing values of external field.  A saturation magnetization of $M_{s} = 8.0 \times 10^5~A/m$, an exchange coupling strength of $C = 13 \times 10^{-12}~J/m$, and damping parameter $\alpha = 0.5$, or a more realistic value of $\alpha = 0.05$ in selected cases, were used as simulation material parameters.  Simulations were run using 4 variations of the same ellipse-pair base model, with dimensions approximating those of the fabricated devices.  The \emph{unbroken} version contains a bridge of finite width ($8-12nm$) between the electrodes; the \emph{point-contact} has a bridge constricting down to a single point; the \emph{nanogap} includes a tapered bridge possessing a roughly $2nm$ gap; and the \emph{ideal} set of leads consists of the ellipses without the constriction.\cite{Scott2016}

\section{RESULTS AND DISCUSSION}
\subsection{Longitudinal Configuration}


\begin{SCfigure*}[][t!]
  \centering
\includegraphics[scale = .3]{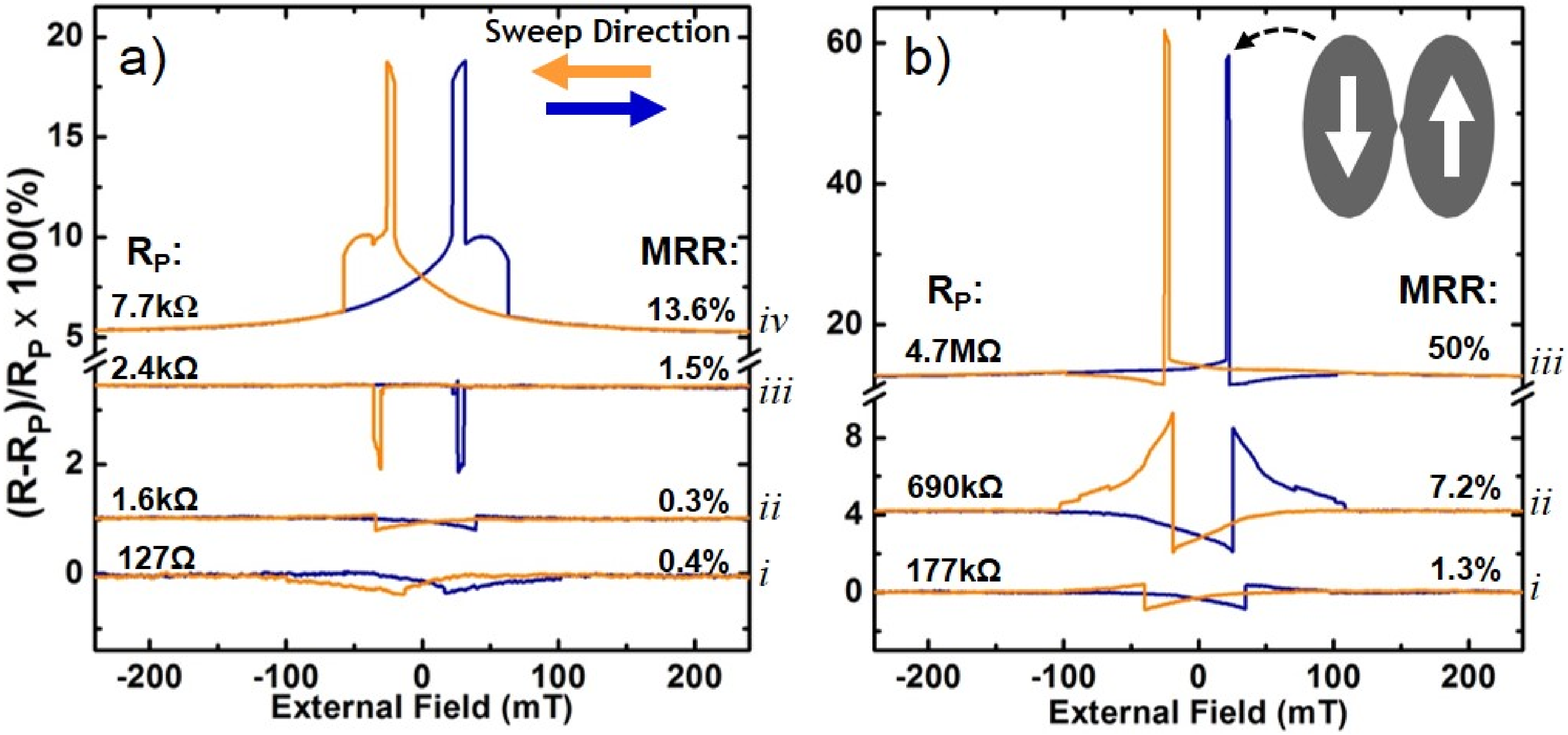}
\vspace{-7mm}
\caption{MR measurements of ellipse-pairs with longitudinal field orientation.  Traces (offset for clarity) exemplify data acquired at different resistance values displaying commonly observed characteristics.  (a) MRR is small and negative for low resistance devices, but often becomes larger as the constriction is reduced in size.  In most, but not all, cases resistance undergoes gradual change prior to an abrupt shift at some critical field, $\textbf{B}_c$.  (b) Larger MRRs are only observed when a suitable tunneling gap is formed.  Inset: When the constriction is not continuous, the high resistance state may correspond to monodomain-like AP alignment of magnetic moments (white arrows) in the electrodes.}
\label{HardData1}
\end{SCfigure*}

\begin{figure}[b!]
\begin{center}
\includegraphics[scale = .22]{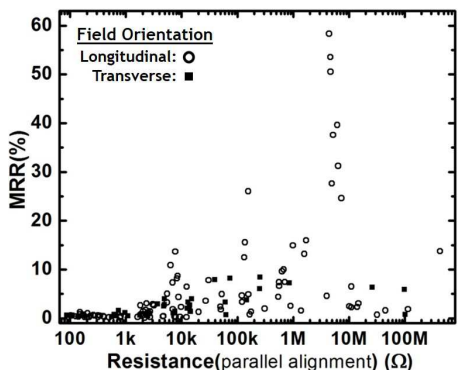}
\end{center}
\vspace{-6mm}
\caption{Absolute value of MRR for all devices tested at various resistance values and corresponding transport regimes.}
\label{MRstats}
\end{figure}

In the high conductance regime, prior to electromigration, as a longitudinally applied field is swept down from $\textbf{B}_s$ ($\simeq1T$), the resistance decreases gradually beginning near $100mT$.  This behavior persists through remanence ($\textbf{B}_{ext}=0T$) until an abrupt change occurs at some critical field, $-\textbf{B}_c$, around $-20mT$ to $-40mT$ (Fig.~\ref{HardData1}a\emph{i},\emph{ii}), leading to MRR values typically no more than $1\%$.  When electromigration is used to reduce the width of the constriction to the point that the resistance is increased to $R\gtrsim{1000}\Omega$, MRR values can increase, but the measured MR behavior generally becomes more varied (Fig.~\ref{HardData1}a\emph{iii},\emph{iv}).  It is only when the resistance is increased by the formation of a sufficiently large tunneling gap in the constriction that we observe increased MRRs (Fig.~\ref{HardData1}b) up to about $60\%$ (Fig.~\ref{MRstats}).

The characteristic features of most $R$ vs. $\textbf{B}_{ext}$ traces are evident for data collected in both sweep directions, but small asymmetries with respect to field inversion may be accounted for if the remanent state is not entirely independent of the direction of saturation because of slight geometric variations between the source and drain electrodes.\cite{Egle2010}.  Resistance changes associated with an enhancement of the saturation field in the constriction region at higher fields are not observed.  The impact of magnetostrictive forces on the measured data is expected to be negligible.  Py$^*$ is a material with low magnetostriction, and the devices are rigidly attached to a Si/SiO$_2$ substrate with no suspended portions.  Furthermore, the measurements are carried out at low temperatures which serves to suppress thermally driven diffusion of atoms near the contact.

\begin{figure*}[t!]
  \centering
\includegraphics[scale = .32]{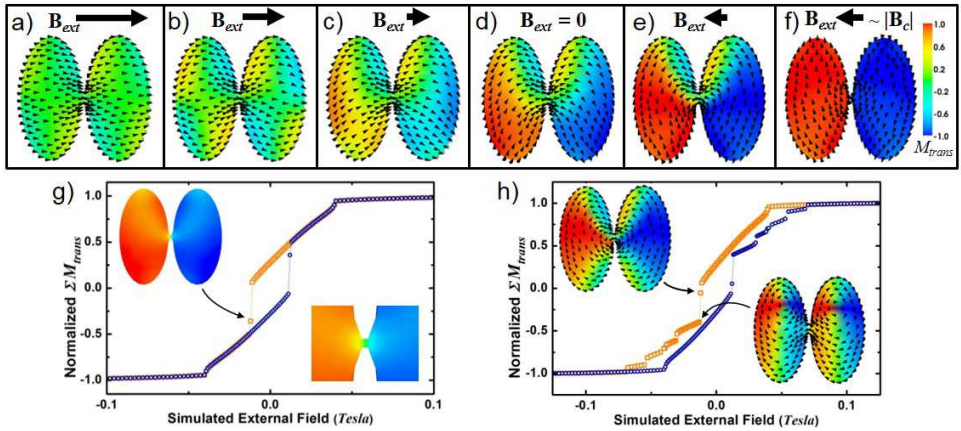}
\vspace{-2mm}
\caption{(a)-(e)  Snapshots of simulated magnetization as $\textbf{B}_{ext}$ is swept down from $\textbf{B}_s$ to $\gtrsim-\textbf{B}_c$.  Colorbar (same for all) indicates normalized local $M_{trans}$ magnitude and black cones denote local magnetization direction.  Simulations show a gradual evolution of the transient spin configuration toward a U (or inverted U) shaped pattern. The impact of the constriction size is primarily exhibited by the accessible intermediate states that may arise as $\textbf{B}_{ext}$ is swept past the critical field.  (f) The monodomain-like AP configuration appears only when using the $\emph{nanogap}$ model.   (g)  Calculated hysteresis loop of normalized $\sum{M_{trans}}$ vs. $\textbf{B}_{ext}$, using the $\emph{unbroken}$ device model.  Insets: snapshot of magnetization distribution near the critical field (top left) showing the bulk of the electrodes develop an approximately AP magnetization distribution.  Zooming in reveals that this does not extend to the constriction, which retains a homogenous longitudinal magnetization (bottom right).  (h) Same as (g), but for the $\emph{point-contact}$ model.  Insets: snapshots before (top left) and after (bottom right) the shift at -$\textbf{B}_{c}$.}
\label{HardSim}
\end{figure*}

These observations are in accord with the simulated magnetization reversal process, which is captured in part by snapshots of the transverse magnetization component magnitudes ($M_{trans}$) and the transient spin configuration at changing values of $\textbf{B}_{ext}$.  The sequence shown in Fig.~\ref{HardSim}a-e indicates that the local magnetization, particularly in the regions along the outer edges of the ellipses, begins to rotate in the plane of the film, reducing the magnetostatic energy, while the constriction region remains uniformly magnetized.  The magnetization of the left and right electrodes rotate counter to one another until a configuration arises wherein the magnetization distribution pattern over the ellipse-pair exhibits an inverted U-shape pattern (Fig.~\ref{HardSim}e)  Throughout this progression local longitudinal component magnitudes, $M_{long}$, decrease and $M_{trans}$ increase, before undergoing an irreversible transition at $-\textbf{B}_{c}$ (Fig.~\ref{HardSim}f).

The MR signature is therefore initially (i.e. prior to $-\textbf{B}_c$) dominated by AMR, strongly influenced by the electrode shape and the stray field generated by the neighboring ellipse\cite{Novosad2003,Yin2011} as the magnetization rotates within each portion of the electrodes.  AMR is manifested as a change in resistivity, $\rho$, when measured as a function of the relative orientation of $M$ and the current density $J$ such that $\rho_{\parallel}~(\vec{M}~{\parallel}~\vec{J})~>~\rho_{\perp}~(\vec{M}~{\perp}~\vec{J})$.  This results from the fact that the angle between $\vec{M}$ and $\vec{J}$ affects orbital overlaps, and thus spin-orbit scattering in the bulk of a ferromagnet, conceivably leading to both abrupt jumps and smooth variations in conductance.\cite{Viret2006,Cuevas1998,Sarau2007}  The simulated evolution from $\textbf{B}_s$ to $\gtrsim-\textbf{B}_c$ (Fig.~\ref{HardSim}a-e) occurs in a roughly equivalent manner for all variations of the ellipse-pair base model, and is thus independent of the constriction or conductance level, yet the measured data does not always show this behavior as the constriction size is reduced.  Although numerical results do not capture all of the nuanced variability evident in the measured data, Fig.~\ref{HardSim}f-h demonstrate that slight variations in the constriction can have a considerable impact on the accessibility and nature of intermediate non-collinear magnetization distribution patterns upon reaching $\pm|\textbf{B}_c|$, which in turn lead to differences in the MR characteristics.  We also note that as the nanocontact constriction is narrowed subtle changes in its geometry will impact the number of conduction channels and their transmission probability, but not identically for both spin orientations.  This may lead to additional, and potentially counteractive, MR contributions due to AMR and domain wall scattering effects.  The net effect can produce MRRs with sample-to-sample fluctuation in size and even sign.\cite{Krzysteczko2008,Egle2010}

As the conductance is reduced below $e^2/h$, AMR will continue to dominate the MR characteristics in the bulk of the ellipses.  However, at the constriction the effect is labeled as AAMR because changes in conduction originate in the atomic nature of the contact.  The reduced dimensions and atomic coordination within the constriction region causes a local variation in the band structure,\cite{Doudin2008} enhancing the spin-orbit coupling which links the shape of the orbitals with the spin direction.\cite{Viret2006}  AAMR can then stem from an associated change in the electronic density of states (DOS) at the nanoconstriction as magnetization is rotated with respect to the current.

The modest MRRs of $\lesssim10\%$ are reasonable for magnetic constrictions down to an atomic point contact.  Although it has been hypothesized that very large MRRs may be anticipated when the local magnetization within an atomic contact changes by $180^{\circ}$ between two neighboring atoms, this scenario is not expected as the exchange energy opposes the anisotropy.\cite{Doudin2008,Egle2010}  This is consistent with our simulations for narrow constrictions and point-contacts (Fig.~\ref{HardSim}g,h), but does not necessarily hold for nanometer-scale tunneling gaps in which a large rotation may exist between spin orientations on the ends of opposing contacts (Fig.~\ref{HardSim}f).

Once a tunneling gap is created, resistance changes can occur due to TMR since the tunneling DOS will be affected by the differing magnetization orientations within the source and drain electrodes.  TMR emerges as a result of a difference between spin-up and spin-down DOS at the Fermi level between two ferromagnets on either side of a tunneling gap.  Tunneling probability is generally reduced and resistance is increased for electrodes with anti-aligned magnetic moments.\cite{Keane2006}  However, if the tunneling barrier is sufficiently small conductance will also be determined by local band structure given by the overlap of wavefunctions of neighboring atoms.  This effect is designated as TAMR and may be rationalized with the same reasoning that applies to AAMR.  Whereas conductance in AAMR is related to the overlap of atomic orbitals, in TAMR it is associated with the overlap of their evanescent tails.  Both effects can lead to enhanced AMR with significant sample to sample variation because both effects are sensitively dependent upon the precise atomic configuration.\cite{Doudin2008,Egle2010}  When the gap is sufficiently large and transport is limited to the tunneling regime we observe MRRs as high as $60\%$.  The TMR and TAMR effects may counterbalance each other to some degree preventing MRRs much greater than this.  Nonetheless, we argue that the increase in MRRs is indicative of AP alignment, in agreement with simulations performed using the \emph{nanogap} model in which an intermediate state is achieved exhibiting electrodes with a nearly homogenous AP magnetization distribution (Fig.~\ref{HardSim}f).


\begin{figure}[t!]
\begin{center}
\includegraphics[scale = .25]{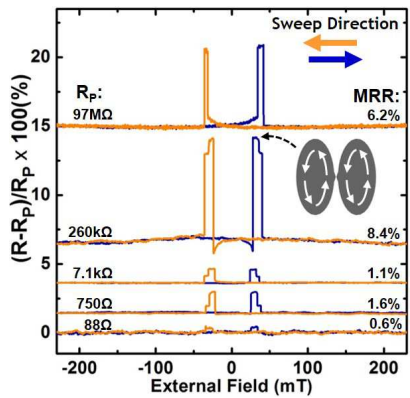}
\end{center}
\vspace{-6mm}
\caption{MR for samples with a transverse field orientation.  Traces offset for clarity.  MRR is typically positive and never observed to exceed $10\%$.  For all conductance regimes the abrupt transition at $\textbf{B}_c$ is rarely preceded by a gradual continuous change, in contrast to the characteristic behavior with longitudinal field, and is most likely associated with intermediate vortex states (Inset).}
\label{EasyData}
\end{figure}

\subsection{Transverse Configuration}

Traces of measured resistance show marked differences when the data is acquired with a transverse field orientation.  Starting from an initially saturated state, as $\textbf{B}_{ext}$ is swept from $\textbf{B}_s$ to $-\textbf{B}_s$, the resistance still undergoes a shift at $-\textbf{B}_c$ ($\approx-20$ to $-40mT$), but gradual changes do not begin until the field has swept past $0.0mT$.  In other words the remnant state is nearly identical to the saturated state.  Resistance jumps at $-\textbf{B}_c$ for a brief range of $\textbf{B}_{ext}$ until it abruptly returns back close to its saturation value.  Like with devices in the longitudinal configuration, the MRRs for devices with transverse field orientation in the high conductance regime are $\lesssim{1}\%$.  When electromigration is used to increase the resistance of these devices, MRR can be observed to increase up to about $8\%$, yet $R$ vs. $\textbf{B}_{ext}$ show mostly similarly shaped traces regardless of the resistance, as seen in Fig.~\ref{EasyData}.

As with the longitudinal configuration, no significant resistance changes are evident at higher fields, and there are no drastic asymmetries with respect to sweep direction.  There are key differences between transverse and longitudinal reversal processes, with the most notable distinction evident when a tunneling gap is created, yet both field orientations lead to many common features including small MRRs in the high conductance regime and abrupt transitions that are confined to low fields ($|\textbf{B}_{ext}|<0.1T$).  Although the range of field over which the resistance changes are apparent is greater for longitudinally applied $\textbf{B}_{ext}$, the typical critical fields are essentially the same for both orientations regardless of constriction size or transport regime.

\begin{SCfigure*}[][b!]
  \centering
\includegraphics[scale = .25]{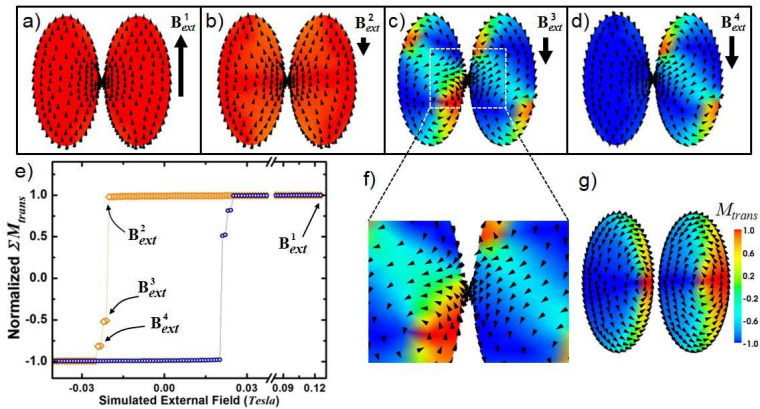}
\vspace{-7mm}
\caption{Snapshots of simulated magnetization using \emph{nanogap} model.  The transient spin configuration exhibits only slight changes as $\textbf{B}_{ext}$ approaches $-\textbf{B}_c$, (a) and (b), consistent with the measured data.  An abrupt reorganization of the magnetization distribution leads to intermediate states (c) and (d).  Colorbar (same for all) indicates local $M_{trans}$ magnitude and black cones denote local magnetization direction.  (e) Calculated hysteresis loop of normalized $\sum{M_{trans}}$ vs. $\textbf{B}_{ext}$, with labeled $\textbf{B}_{ext}$ values corresponding to (a)-(d) snapshots.  (f) Zoom in of (c) showing a nearly AP magnetization configuration localized at the constriction.  (g) Intermediate state at $\textbf{B}^3_{ext}$ when the simulation is repeated using the \emph{ideal} model.}
\label{EasySim}
\end{SCfigure*}

The observations are in agreement with micromagnetic simulations which indicate that the local magnetization distribution within the electrodes changes little prior to reaching the critical field, remaining predominantly directed along the easy axis.  Fig.~\ref{EasySim}a,b shows that as $\textbf{B}_{ext}$ is swept from $\textbf{B}_s$ to $\gtrsim-\textbf{B}_c$ the magnetization distribution within the electrodes begins to rotate only slightly, primarily along the inner edges of the ellipses, leading to a small reduction in $M_{trans}$ and increase in $M_{long}$.  This evolution is seen for all variations of the ellipse-pair base model with transverse field orientation and is consistent with resistance changes dominated by conventional AMR.

When $-\textbf{B}_c$ is reached, the simulations using the \emph{nanogap} model demonstrate a few possibilities for the resulting local spin configurations that may occur as the demagnetizing fields within each particle cease to be counter balanced by the external and exchange fields.  Intermediate states may exhibit displaced vortices combined with buckling modes in both ellipses (Fig.~\ref{EasySim}c) or in only one ellipse while the other has a roughly homogenous magnetization (Fig.~\ref{EasySim}d).  The corresponding hysteresis loop of normalized $\sum{M_{trans}}$ is shown in Fig.~\ref{EasySim}e.  Intermediate states with flux closure patterns can lead to a local AP magnetization distribution at the site of the tunneling gap or atomic contact, as seen more clearly in Fig.~\ref{EasySim}f.  When the \emph{ideal} model is used numerical results show accessible intermediate states with a magnetization distribution in which a single displaced vortex has nucleated in each ellipse (Fig.~\ref{EasySim}g).  This demonstrates that the constriction is significant to the model simulation, but does not fundamentally change the qualitative characteristics of the magnetization reversal.  Thus the abrupt increase in $R$ seen in all traces of Fig.~\ref{EasyData} may be associated with various intermediate vortex states.

Furthermore, when electromigration is used to increase the resistance of these devices, MR signatures may similarly be expected to result from AMR and AAMR until a sufficiently large tunneling gap allows for transport influenced by TAMR and TMR.  However, for devices with a transverse field orientation, $R$ vs. $\textbf{B}_{ext}$ traces exhibit less variability within all resistance ranges tested.  And although both AAMR and TAMR lead to sample-to-sample MRRs with a significant scatter in magnitude, MRR values remain below $10\%$.  We surmise that the limitation on these values stems from the use of the transverse field orientation which is constrained to producing intermediate state configurations like vortex patterns, whereas the larger MRR values only found here when using a longitudinal field orientation in the tunneling regime are indicative of the desired AP alignment.

\section{CONCLUSION}

We demonstrated that ferromagnetic electrodes with nanometer separation capable of contacting an individual molecule may be switched between P and AP magnetization configurations by exploiting the magnetization reversal process with appropriately shaped electrodes and proper choice of external field orientation.  A pair of planar ellipses elongated along the axis perpendicular to the constriction between them is identified as a suitable design for achieving this control when the external field modulating the reversal process is applied along their shared short axis.  As the constriction size diminishes, MR traces can show a spectrum of possible variations in accord with AMR, AAMR, and TAMR mechanisms.  Once a sufficient tunneling gap is created, MR will be primarily dictated by TMR, but counterbalanced by TAMR.  Micromagnetic simulations were qualitatively consistent with measured data.  The findings demonstrate an important means by which control can be exerted over particular magnetic properties of nanostructured materials.  This work may also spark renewed interest in experimental efforts with single molecule transistors incorporating magnetic materials by providing numerical results against which future MR data may be benchmarked.

Thanks to Rose Kopf and Al Tate for help with fabrication and Gerardo Gamez for useful discussions.

\vspace{-3mm}

\bibliographystyle{apsrev}

\bibliography{PyMfabA_noURL}

\end{document}